\documentclass{cpbtex}
\usepackage{amssymb}
\usepackage{graphicx}

\begin{document}
\begin{CJK*}{GBK}{song}

\title{Thermoelectric Effects and Topological Insulators\thanks{Project supported by the National Thousand-Young-Talents Program and Tsinghua University Initiative Scientific Research Program.}}
\author{Yong Xu$^{1,2,3}$\thanks{Corresponding author. Email: yongxu@mail.tsinghua.edu.cn} \\
$^1${State Key Laboratory of Low Dimensional Quantum Physics, Department of Physics,}\\
{Tsinghua University, Beijing 100084, People's Republic of China}\\
$^2${Collaborative Innovation Center of Quantum Matter,}\\
{Beijing 100084, People's Republic of China}\\
$^3${RIKEN Center for Emergent Matter Science (CEMS), Wako, Saitama 351-0198, Japan}}
\maketitle
%\date{\today}

\begin{abstract}
The recent discovery of topological insulator (TI) offers new opportunities for the development of thermoelectrics, because many TIs (like Bi$_2$Te$_3$) are excellent thermoelectric (TE) materials. In this review, we will first describe the general TE properties of TIs and show that the coexistence of the bulk and boundary states in TIs introduces unusual TE properties, including strong size effects and anomalous Seebeck effect. Importantly, the TE figure of merit $zT$ of TIs is no longer an intrinsic property, but depends strongly on the geometric size. The geometric parameters of two-dimensional TIs can be tuned to enhance $zT$ to be significantly greater than 1. Then a few proof-of-principle experiments on three-dimensional TIs will be discussed, which observed unconventional TE phenomena that are closely related to the topological nature of the materials. However, current experiments indicate that the metallic surface states, if their advantage of high mobility is not fully utilized, would be detrimental to TE performance. Finally we provide an outlook for future work on topological materials, which offers great possibilities to discover exotic TE effects and may lead to significant breakthroughs in improving $zT$.
\end{abstract}


%73.50.Lw Thermoelectric effects
%72.20.Pa Thermoelectric and thermomagnetic effects
%71.90.+q Other topics in electronic structure (restricted to new topics in section 71)
%73.23.-b Electronic transport in mesoscopic systems

\section{Introduction}
Thermoelectric effect is a challenging research topic that has been studied for nearly two hundred years. The recent discovery of topological insulator, on the other hand, is one of the most important progresses in condensed matter physics and materials science. The intersection of these two research topics would open doors for novel phenomena and new opportunities that are not available in conventional systems. In this review, we will first present a general introduction on thermoelectrics, including the research background, main challenges and important progresses of the field. Then we will discuss unique/novel thermoelectric properties of topological insulators mainly from theoretical predictions, like the strong size effect and anomalous Seebeck effect. After that we will show some recent proof-of-principle experiments on thin films of three-dimensional topological insulators. Finally we will give an outlook for future thermoelectric research.

\subsection{Research background of thermoelectrics}
Future energy use must be sustainable so as to increase economic prosperity with no environmental sacrifice. One of promising solutions is to use thermoelectric (TE) materials for power generation and refrigeration. Currently about 90\% world's power ($\sim$10 terawatts) is generated by heat engines based on fossil fuel combustion. The corresponding energy conversion efficiency is typically 30\%-40\%, leading to $\sim$15 terawatts waste heat into environment\ucite{hochbaum2008}. The huge amount of waste heat, if being recycled, would greatly ease the growing energy and environmental crisis nowadays. This stimulates intensive research interest on TE effects that are interconversion of heat and electricity. The TE effects includes two types of effects: the Seebeck effect and Peltier effect. The former effect converts a temperature gradient into an electric voltage, useful for waste heat recovery. The later effect generates a heat current by applying an electric current, useful for cooling and heating.

Distinguished from conventional heat engines (e.g. gas turbines), TE devices are solid state energy convertors using electrons as working liquid\ucite{bell2008}. They have no moving parts, thus are very silent and reliable. Their energy efficiency does not decrease with device miniaturization, advantageous for power supply of nanodevices. Also, TE power is highly scalable, covering from a few milliwatts to hundreds of megawatts. Benefiting from these advantages, TE devices have been widely used in various applications, including space exploration (e.g. radioisotope TE generator of Mars Rover), military use (e.g. infrared night vision with TE cooling), everyday life use (e.g. heat-electricity cogeneration, portable refrigerator and thermally-controlled car seat) and waste heat recovery. However, the currently used TE devices, though having attractive features, are limited by their low efficiency. Further performance improvement, even if slight, would greatly promote large-scale applications.

The energy conversion efficiency of TE devices is determined by the figure of merit $zT$ of TE materials. $zT$ is a dimensionless quantity defined as
\begin{eqnarray}
zT = \frac{\sigma S^2 T}{\kappa_e + \kappa_l},
\end{eqnarray}
where $\sigma$ is the electrical conductivity, $S$ is the Seebeck coefficient, $T$ is the temperature, and the thermal conductivity $\kappa = \kappa_e + \kappa_l$ includes contributions from electrons ($\kappa_e$) and lattice vibrations ($\kappa_l$). For a high-performance TE material, it should have large $S$ so as to effectively convert temperature differences into electric voltages. It is also required to have large $\sigma$ and small $\kappa$ so that the energy loss induced by Joule heating and heat diffusion is minimized. The larger $zT$, the higher the TE efficiency.

\subsection{Main challenges and important progresses of thermoelectrics}
However, improving $zT$ is quite challenging, because $zT$ is a combination of conflicting physical quantities. Generally, introducing defects/disorders into the materials would scatter phonons and decrease $\kappa$, which would also scatter electrons and suppress $\sigma$; increasing charge carrier concentration would enhance $\sigma$ but lower $S$, as illustrated in Fig.~1(a)~\ucite{snyder2008}. A compromise between $\kappa$, $\sigma$ and $S$, realized by tuning defect/disorder scatterings and charge carrier concentration, is usually required to optimize $zT$.

The study of new materials could greatly promote TE developments. If looking back to the history of thermoelectrics~\ucite{ioffe1957,rowe1995,goldsmid2010}, one will soon find that previously significant progresses of TE research were always related to the application of new materials. At the very beginning TE effects were discovered in metals. However, the Seebeck effect in metals is very weak. TE effects were thought to be nearly useless, which were only applied for thermocouples. After the discovery and wide application of semiconductor materials, people started to explore thermoelectric properties of semiconductors, and soon found that TE performance of semiconductors is much better than that of metals. For metallic systems, the electronic states below the Fermi level $E_F$ have a positive contribution to $S$; while the electronic states above $E_F$ have an opposite contribution to $S$. The two contributions compensate with each other, leading to small $S$. For semiconductors, the materials can be either N-type or P-type controlled by doping, with either electrons or holes as the main charge carriers. Since the compensation between electrons and holes is avoided, strong Seebeck effect is obtained in semiconductors. TE properties of semiconductors thus attracted intensive research interest.

The fundamental theory of thermoelectrics has become gradually mature till 1950-1960. For semiconducting materials, there are various guiding principles of improving $zT$~\ucite{ioffe1957,rowe1995,goldsmid2010}, through tuning charge carrier concentration, band gap, the type of chemical bonding and chemical composition, which are briefly described here: (1) Low doping concentration gives large $S$ but small $\sigma$. Increasing doping concentration enhances $\sigma$ but suppresses $S$. For most semiconductors, $zT$ is optimized at a charge carrier concentration of around $10^{19}$-$10^{20}$ cm$^{-3}$, as illustrated in Fig.~1(a)~\ucite{snyder2008}. (2) Previous studies indicate that the power factor $\sigma S^2$ gets maximized when the band gap is around 6-10 $k_B T$, where $k_B$ is the Boltzmann constant. This corresponds to a band gap of 0.1-0.3 eV for the room temperature. (3) Ionic compounds usually have strong electron-phonon interactions and thus small charge carrier mobility $\mu$. In comparison, covalent compounds typically gives relatively larger $\mu$, advantageous for high $zT$. (4) Materials composed of heavy elements give low phonon frequency, small group velocity and thus low thermal conductivity of lattice vibrations, favored by thermoelectrics. Based on these guiding principles, people found the state-of-the-art room-temperature thermoelectric materials~\ucite{goldsmid1954}, including Bi$_2$Te$_3$, Sb$_2$Te$_3$, Bi$_2$Se$_3$, etc. From metals to semiconductors, $zT$ has been increased considerably from $zT \ll 1$ to $zT \sim 1$. Further enhancement of $zT$, however, is quite difficult mainly caused by the entanglement between materials' thermal and electronic transport properties. An effective decoupling between thermal and electronic transport is crucial to realize high $zT$.

One of most important concept of TE research is phonon-glass-electron-crystal (PGEC)~\ucite{rowe1995}. In PGEC systems, the transport of phonons is strongly scattered like in a disordered glass, while the transport of electrons experiences little scattering like in a perfect crystal. This kind of systems are promising for thermoelectrics. New material structures and physical mechanisms are required to design PGEC systems. One widely used strategy is through alloying (or site substitution) with isoelectronic elements (like Bi$_2$Te$_3$ with Sb$_2$Te$_3$ or Bi$_2$Se$_3$)~\ucite{snyder2008}. In such kind of alloys, a crystalline electron structure is preserved for electrons, but a disordered phonon structure is induced by the large mass contrast. As an example, P-type Bi$_x$Sb$_{2-x}$Te$_3$ and N-type Bi$_2$Se$_{y}$Te$_{3-y}$ belong to the best room-temperature TE materials. Another strategy is to introduce disorders into complex lattice structures~\ucite{snyder2008}. For instance, as shown in Fig.~1(b) PGEC systems can be effectively constructed in the skutterudite CoSb$_3$ structure by adding defect atoms into the blue void space. The disorder strongly scatter phonons, but have little influence on electrons, whose transport path is outside the void space. Based on the feature that the transport paths of electrons and phonons are separated in real space, one can selectively scatter phonons keeping electrons little affected by controlling the distribution of defect atoms. The kind of ``modulation doping'' or substructure approach can also be applied to layered materials, like Na$_x$CoO$_2$. As shown in Fig.~1(c), the interlayer region is filled with randomly distributed sodium atoms, i.e. ``phonon glass''; the intralayer region as the transport path of electrons (shaded red) is ordered, i.e. ``electron crystal''.

After the discovery of Bi$_2$Te$_3$-series alloy materials with room-temperature $zT \sim 1$, intensive research attempted to find better TE materials, but very few breakthroughs were achieved for a long time. During 1960-1990, TE research of semiconductors had faced a bottleneck. People gradually realized that it is extremely difficult to realize $zT > 1$ in traditional materials. In 1990s, nanoscience and nanotechnology come to birth and rise very soon. TE research started to study nanomaterials, which support characteristics significantly distinct from their bulk counterparts, caused by the profound quantum confinement effects. The concept of improving $zT$ by nanostructuring is first proposed by Hicks and Dresselhaus in 1993~\ucite{hicks1993,hicks1993_2}. They proposed that in low-dimensional systems, including two-dimensional quantum wells and one-dimensional nanowires, the quantum confinement effects could largely modify the electronic band structure and thus improve the power factor. Nanostructuring materials could also help lower thermal conductivity of lattice vibrations. Therefore, $zT$ could be considerably enhanced in nanomaterials.

Although quantum confinement effects were predicted to enhance $zT$ majorly by increasing power factor ~\ucite{hicks1993,hicks1993_2}, in experiment the improvement of $zT$  in nanostructures mainly benefits from the decrease of lattice thermal conductivity. One representative example is silicon nanowire. Bulk silicon has a very low $zT$ of about 0.01 at room temperature. In 2008 two experiments discovered that $zT$ can be dramatically enhanced in silicon nanowires to $\sim$1, about 100 times higher than their bulk counterpart~\ucite{hochbaum2008,bell2008}. The underlying mechanism is that electrons and phonons have distinctly different characteristic lengths (i.e. wavelength and mean free path). The characteristic length of phonons is much longer than that of electrons. As a result, in thin silicon nanowires of diameters around tens of nanometers, the rough boundary strongly scatters long-wavelength phonons and decreases the lattice thermal conductivity considerably by 100 times; while charge carriers that are mainly distributed within the nanowire are almost unaffected by the rough boundary. Therefore, the decrease of thermal conductivity together with the preservation of power factor increase $zT$ by 100 times in thin silicon nanowires. In fact, it is a very active research field to develop silicon-based thermoelectric materials with nanostructures, due to the technology importance of silicon and the compatibility with the current semiconductor industry. The reduction of lattice thermal conductivity, which is the main reason for the enhanced $zT$, is induced by two typical mechanisms: (i) incoherent mechanisms that reduce thermal conductivity by scatterings, for instance via interface/surface~\ucite{chen2010,chen2011} or defect/impurity~\ucite{chen2009,shi2010}, and (ii) coherent mechanisms that modulate the phonon band structure and group velocity, for instance via nanomesh structure~\ucite{yu2010,tang2010} or core-shell structure~\ucite{chen2011_2,chen2012,wingert2011}.

There is another class of materials promising for thermoelectrics, which have unusually low intrinsic thermal conductivity. Typically the lattice thermal conductivity decreases with increasing temperature in the form of $T^{-1}$, due to the increased phonon-phonon scattering and decreased phonon mean free path. However, an experiment in 2008 observed that the intrinsic lattice thermal conductivity of the cubic I-V-VI$_2$ semiconductors (like AgSbTe$_2$ and AgBiSe$_2$) is essentially temperature independent from 80 to 300 K, keeping at an extraordinarily low value of  $<$1.0 W/m-K, close to the lower amorphous limit of solids (Fig.~2)~\ucite{morelli2008}. In such kind of materials, the anharmonic interactions and the associated phonon-phonon scattering are extremely strong, inducing a very short phonon mean free path at low temperatures, as short as the atomic distance. The increase of temperature cannot further decrease the phonon mean free path any more. Therefore, the thermal conductivity is very low and nearly temperature independent. Such kind of materials can simultaneously have a crystalline electronic structure together with an ultralow intrinsic thermal conductivity, which behave naturally like PGEC systems. The recently investigated material SnSe belongs to this material class. An experiment in 2014 found a record high $zT$ value of $\sim$2.6 at 923 K in the high-temperature phase of single crystal SnSe~\ucite{zhao2014}. Later in 2016 the same experimental group tuned the charge carrier concentration by controlling dopants and  observed $zT$ values of 0.7-2.0 from 300 to 773 K in the low-temperature phase of SnSe~\ucite{zhao2016}. We expect more high-performance TE material candidates to be found in this material class in the near future.

\subsection{Topological insulators as thermoelectric materials}
The discovery of topological insulators (TIs)~\ucite{kane2005,bernevig2006,konig2007} is one of most important progresses in condensed matter physics and material science in the recent years~\ucite{qi2010,hasan2010,qi2011}. TIs are new states of quantum matter, characterized by insulating bulk states and metallic boundary states. The gapless boundary states are helical with spin-momentum locking and topologically protected against backscattering by time reversal symmetry. TIs can be classified into three-dimensional (3D) TIs (e.g. Bi$_2$Te$_3$, Sb$_2$Te$_3$, Bi$_2$Se$_3$) with two-dimensional (2D) metallic surface states~\ucite{Zhang2009,Chen2009,Xia2009} and 2D TIs (e.g. HgTe/CdTe quantum well) with one-dimensional (1D) metallic edge states~\ucite{bernevig2006,konig2007}. The nontrivial surface or edge states are featured by the unique Dirac-like linear dispersion, spin polarization and topological protection, which can find important applications in various fields, including electronics, spintronics, quantum information and so on~\ucite{qi2010,hasan2010,qi2011}. Moreover, TIs provides an ideal platform to explore emerging new physics, like Majorana Fermion, magnetic monopole and quantum anomalous Hall (QAH) effect~\ucite{qi2010,hasan2010,qi2011,wang2013,zhang2014,zhang2014_2,zhao2015}.

The recent discovery of TIs also offers new opportunities for the development of thermoelectrics. In fact, many TIs are excellent TE materials, which include the second generation TI Bi$_x$Sb$_{1-x}$, the third generation TI Bi$_2$Te$_3$, Sb$_2$Te$_3$ and Bi$_2$Se$_3$, and some other TIs like Bi$_2$Te$_2$Se, Bi$_2$Te$_2$S, Tl(Bi,Sb)(Te,Se,S)$_2$, Ge1Bi$_4$Te$_7$, Ge$_2$Bi$_2$Te$_5$, GeBi$_2$Te$_4$, PbBi$_2$Se$_4$, PbSb$_2$Te$_4$, ZrTe$_5$, HfTe$_5$, ternary Heusler compounds, filled skutterudite and so on. Although these TE materials have been intensively studied, their nontrivial boundary states have not been considered in early TE research, which may play an important role in improving TE performance. What is the underlying relation between topology and TE effects? Is there any fundamentally new concept for using TIs as TE materials? How to use the novel TI states for TE effects? These are important problems to be explored.

\section{Thermoelectric effects in topological insulators}
\subsection{Essential features of topological insulators}
In fact, TIs and TE materials share similar material traits including heavy elements and narrow band gap. So that for TIs, they have strong spin-orbit coupling (SOC) to induce a band inversion; for TE materials they have low thermal conductivity and large power factor. This could partially explain the fact that many TIs are excellent TE materials.

In contrast to conventional insulators or semiconductors, TIs are unique in their electronic properties. They have insulating bulk states as well as metallic boundary states that are topologically protected to appear within the bulk band gap. The bulk and boundary states are different on various aspects: (1) The bulk states distribute homogeneously within the material interior, while the boundary states distribute mainly on the material surfaces/edges. (2) The bulk states have a band gap and are occupied by massive Fermions, while the boundary states are gapless with Dirac-like linear dispersions and occupied by massless Fermions. (3) The bulk states are easily scattered by disorders and defects, while the boundary states are immune to backscattering of nonmagnetic defects as protected by time reversal symmetry, and thus have much better transport ability than the bulk states. The two distinctly different kinds of states coexist and jointly contribute to TE transport. Therefore, if tuning their relative contribution to TE transport, the overall TE properties would change. This new degree of freedom that can be controlled by varying geometric size, applying external field, inducing topological phase transition, etc, makes TE properties of TIs interesting.

\subsection{Size dependent thermoelectric properties in topological insulators}
The key challenge of TE research is to improve $zT$. However, we find that the typical definition of $zT$ has to be changed fundamentally in TIs~\ucite{xu2014}. The conductivity quantities $\sigma$ and $\kappa$ are used in the typical definition of $zT$, which implicitly assumes that $zT$ is an intrinsic material property, independent of the geometric size. This basic assumption is not true for TIs, due to the coexistence of two types of states with different geometric size dependence.

To describe the size dependence of TE properties, we derive a general definition of $zT$ from thermodynamics:
\begin{eqnarray}
zT = \frac{G S^2 T}{K},
\end{eqnarray}
where $G$ is the electrical conductance and $K$ is the thermal conductance. Different from in the typical definition,  conductance quantities $G$ and $K$ that vary with geometric sizes (i.e. the cross sectional area $A$ and transport length $L$) are used instead of conductivity quantities. The two types of definitions are equivalent to each other, providing the following two conditions are satisfied: (i) the electrical conductance satisfies the Ohm's scaling law (i.e., $G \propto A/L$) and the thermal conductance satisfies the Fourier's scaling law (i.e., $K \propto A/L$). This condition implies $G/K = \sigma / \kappa$. (ii) $S$ is independent of the geometric size, or $S$ does not changes with $A/L$. $zT$ would be size independent if and only if both conditions are satisfied.

The above derivation tells us that in general there are two mechanisms inducing size dependence of $zT$. One is the failure of the Ohm's or Fourier's scaling law. The other is the dependence of $S$ on the geometric size. In TIs both mechanisms take effects: (i) When decreasing $A$, the contribution of the bulk states to TE transport gets smaller, while the contribution of the boundary states remains unchanged. The relative contribution of the boundary states thus increases. On the other hand, the transport of the boundary states, if not diffusive, would decay with $L$ much slower than that of the bulk states. When increasing $L$, the relative contribution of the boundary states to TE transport would get enhanced. The different dependence on $A$ and $L$ between the boundary and bulk states results in a feature that $G$ is not proportional to $A/L$, suggesting that the Ohm' scaling never works for TIs (especially in nanostructures). (ii) The Seebeck effect is contributed by the boundary and bulk states (denoted by subscripts ``1'' and ``2'', respectively), $S = (G_1 S_1 + G_2 S_2)/ (G_1 + G_2)$. Even if $S_1$ and $S_2$ are both independent of the geometric size. Their relative contribution to the total $S$, determined by the ratio of electrical conductance $G_1 / G_2$, varies with the geometric size. The total $S$ thus changes with the geometric size. Therefore, $zT$ of TIs are expected to be size dependent, which suggests a simple way of improving zT by optimizing the geometry size~\ucite{xu2014}.

Here we will mainly discuss TE properties of 2D TIs. Similar discussion can be made for 3D TIs. 2D TIs support the intriguing quantum spin Hall (QSH) effect, characterized by the gapless edge states with Dirac-like linear dispersions [Fig.~3(a)]. The nontrival edge states are helical, whose spin and momentum are locked, with the up and down spins moving in the opposite directions. The edge states are immune to nonmagnetic defects, leading to dissipationless transport that is advantageous for enhancing TE performance. However, due to the gapless feature, the nontrivial edge states are thought to have very small $S$, disadvantageous for thermoelectrics. For enhancing $S$, early work proposed to open a hybridization band gap between boundary states~\ucite{ghaemi2010}, which unfortunately would degrade the transport ability of the boundary states. Instead, we are trying to use the intrinsic properties of the boundary states to get strong Seebeck effect with their excellent transport ability preserved.

TE quantities of electrons and phonons can be described by their transmission functions~\ucite{xu2009,xu2010,zhu2012,chen2013,huang2013,li2014,zou2015,chen2015}. When the electronic transmission function $\overline {\cal T} (E)$ is a smooth function and the temperature is low, the Sommerfield expansion of $S$ gives~\ucite{paulsson2003}
\begin{equation}
 S = {\left. { - \frac{{{\pi ^2}k_{\rm{B}}^2T}}{{3e}}\frac{{\partial \ln [\overline {\cal T} (E)]}}{{\partial E}}} \right|_{E = {E_{\rm{F}}}}},
\end{equation}
where $\overline {\cal T} (E) = M(E) {\cal T} (E)$. The distribution of conduction channel $M(E)$ corresponds to the number of conduction channel at a given energy $E$, which is proportional to the group velocity (along the transport direction) times the density of states. ${\cal T} (E)$ is the transmission probability determined by the mean free path $\lambda(E)$ or the scattering time $\tau(E)$. ${\cal T} (E) = \lambda(E) / [\lambda(E) + L]$ is satisfied for ballistic-diffusive transport when the coherent effect of successive scatterings is not important~\ucite{xu2008,xu2014small}. $M(E)$ is determined by the electronic band structure and ${\cal T} (E)$ is related to scattering. To get a large $S$, $\overline {\cal T} (E)$ is required to have a large slope around $E_F$. This can be realized by two mechanisms: (i) Increase the energy dependence of $M(E)$ through, for instance, opening an energy gap or inducing a peak in the density of states. (ii) Increase the energy dependence of $\lambda(E)$ or $\tau(E)$. The second mechanism is usually neglected in most theoretical research that assumes constant mean free path or constant scattering time for simplicity. As we will show, the later mechanism plays a crucial role in TIs.

To describe $\tau(E)$ of the nontrivial edge states, two important facts have to be considered: (i) when $E_F$ is located within the bulk-gap region, the edge states are protected against backscattering by time reversal symmetry, giving rise to  very large $\tau$. (ii) When $E_F$ is located outside the bulk-gap region, the edge-bulk interactions make backscattering possible, which largely decreases $\tau$. Transport calculations usually use the constant scattering time approximation. The approximation would be well suitable for 1D linear bands and 2D parabolic bands that have constant density of states, if assuming the scattering rate $1/\tau$ is proportional to the density of states (like for electron-phonon scattering). For the nontrivial edge states, we proposed a dual scattering time (DST) model that assumes two different constant scattering times $\tau_1$ and $\tau_2$ for the edge states within and outside the bulk-gap region, with $\tau_1 \gg \tau_2$ as illustrated in Fig.~3(a)~\ucite{xu2014}. When $E_F$ is close to the bulk band edge, $\tau(E)$ of edge states shows a strong energy dependence, which is able to induce a large $S$ as demonstrated in Fig.~3(b). In addition, $S$ of the edge states shows an unusual sign. When $E_F = 0$ (referenced to the bottom of bulk conduction band), the Fermi level is slightly above the Dirac point [Fig.~3(a)], the type of carrier carriers of the edge states is N-type as determined from the Hall effect. While the calculated $S$ is positive [Fig.~3(b)], showing P-type Seebeck effect. This can be understood as follows. It is well known that the electronic states below and above $E_F$ have oppositive contributions to $S$. When the states above $E_F$ have a dominated contribution, N-type Seebeck effect will be obtained with $S < 0$; otherwise P-type Seebeck effect is obtained with $S > 0$. In the present system with $E_F \sim 0$, the edge states above $E_F$ can be strongly scattered by defects/disorders due to the edge-bulk interactions. Thus the Seebeck effect is mainly contributed by the edge states below $E_F$, giving $S > 0$. Therefore, the anomalous sign of the Seebeck effect (opposite to the sign of the Hall effect) for the nontrivial edge states is induced by the unique energy dependence of $\tau(E)$ caused by the edge-bulk interactions.

A general discussion on TE properties of TIs has been made~\ucite{xu2014} and some related discussions can be found in elsewhere~\ucite{ghaemi2010,takahashi2010,murakami2011,takahashi2012,chang2014}. Next we show an example study of a specific material, fluorinated stanene~\ucite{xu2013}. Stanene is an atomic layer of tin atoms arranged in a buckled honeycomb lattice, which is a 2D TI similar as graphene. However, in contrast to graphene, stanene is composed of heavy element and has a buckled structure, thus its effective spin-orbit coupling is much stronger, leading to a bulk band gap five orders of magnitude larger ($\sim$0.1 eV of stanene vs. $\sim$$\mu$eV of graphene). Interestingly, the decorated stanenes (e.g. fluorinated stanene), constructed by chemically saturating the $p_z$ orbitals of tin atoms, are also 2D TIs, and their nontrivial band gap can be even larger ($\sim$0.3 eV), supporting room-temperature QSH effect~\ucite{xu2013}. For nanoribbon structures of fluorinated stanene, we calculated the room-temperature $zT$ as a function of the nanoribbon width $W$ and the transport length $L$ [see Fig.~3(c)]. The results indicate that $zT$ is strongly size dependent. When decreasing $W$ and increasing $L$, the contribution of edge states to TE transport gradually increases, leading to an enhanced $zT$. There exists a sizable region of geometric parameters, where $zT > 3$. TE performance of TIs can be significantly improved by controlling the geometric size for maximizing the contribution of the edge states to TE transport. This approach is quite simple and effective. As an evidence, a recent first-principles study reported very high $zT\sim 3$ at $T = 40$K in nanoribbon structures of graphene-based 2D TI~\ucite{chang2014}.

People may wonder why such a high $zT$ can be realized in 2D TIs. The underlying reason is simple. For nanoribbon structures of 2D TIs, when decreasing the width of nanoribbons, the thermal conductance of lattice vibrations (or phonons) gets smaller, while the contribution of edge states does not change. In addition, when introducing defects and disorders into the material, phonons will be strongly scattered, while the transport of the edge states will be little affected. Therefore, due to the existence of the nontrivial edge states, the contribution of electrons and phonons can be effectively decoupled. In this sense, 2D TIs can be viewed as PGEC systems that support high $zT$. Note that our discussion mainly focuses on 2D TIs. TE properties of 3D TIs has been studied extensively~\cite{venkatasubramanian2001, huang2008, hor2009, ghaemi2010, zahid2010, yu2014, rittweger2014, osterhage2014, gooth2014, yu2015, hinsche2015, shi2015_2, liu2016, liang2016}, which will not be elaborated here.

Based on the unique electron properties of TIs, in particularly the coexistence of the boundary and bulk states together with the boundary-bulk interactions, a few theoretical predictions have been made~\ucite{xu2014}: (i) Thermoelectric properties of TIs can show strong size effects. (ii) TIs can give anomalous Seebeck effect, whose sign is opposite to that of the Hall effect. Similar features are expected to appear in other topological materials, like topological crystalline insulators and QAH insulators.

\subsection{Proof-of-principle experiments}
Next we briefly review some recent proof-of-principle experiments. In collaboration with Prof. Yayu Wang, Prof. Ke He and Prof. Qi-Kun Xue, we experimentally investigated thermoelectric properties of 3D TIs and found a few unusual TE phenomena, including the opposite signs between the Seebeck and Hall effects at all the measured temperatures (up to 300 K)~\ucite{zhang2015} and the strong size effects of TE quantities~\ucite{guo2016}. These novel TE phenomena are closely related to the topological features of the materials, which directly or indirectly proves the above theoretical predictions~\ucite{xu2014}.

In the study of semiconductors, people usually apply either the Hall effect or the Seebeck effect to measure the type of charge carriers. N-type charge carriers give negative Hall coefficient $R_H$ and negative $S$. P-type charge carriers give positive $R_H$ and $S$. The sign of $R_H$ and $S$ is typically the same, implying that the Hall and Seebeck measurements give consistent results on the type of charge carriers. However, recently we observed opposite signs between $R_H$ and $S$ in 3D TIs (Bi$_{1-x}$Sb$_x$)$_2$Te$_3$ ($x=0.9$) thin films of thickness 5 quintuple layers (QLs) ($\sim$5 nm)~\ucite{zhang2015}. The material exhibits N-type Hall effect and P-type Seebeck effect, and the feature remains unchanged when varying the temperature [Fig.~4(a)]. Such kind of unconventional phenomenon, as far as we know, has never been reported before.

To understand the observed effects, we measured the electronic band structure of the TI thin films by angular resolved photoemission spectroscopy (ARPES). The measurement indicates that $E_F$ of the material sample with ($x=0.9$) is slightly above the top of the bulk valence band and above the Dirac point of the surface bands, suggesting that the bulk states are P-type and the surface states are N-type, as illustrated in Fig.~4(b). The Seebeck effect and the Hall effect are contributed by two types of charge carriers:
\begin{eqnarray}
S = \frac{-\sigma_s |S_s| + \sigma_b S_b}{\sigma_s + \sigma_b}, \ \  R_H = \frac{-\sigma_s \mu_s + \sigma_b \mu_b}{(\sigma_s + \sigma_b)^2},
\end{eqnarray}
where the subscripts ``s'' and ``b'' denote surface and bulk states, respectively, $\mu$ is the mobility of charge carriers, $S_s < 0$ and $S_b > 0$. The key parameter is the conductivity ratio between the bulk and surface states, $\sigma_b / \sigma_s$. As shown in Fig.~4(c), when $\sigma_b / \sigma_s$ is very small, the surface states dominate both the Seebeck and Hall effects, and thus $S$ and $R_H$ are N-type, showing negative signs (Region I). In the opposite limit of very large $\sigma_b / \sigma_s$, both $S$ and $R_H$ are dominated by the bulk states and thus are P-type, showing positive signs (Region III). In the intermediate region, when $|S_s / S_b| < \sigma_b / \sigma_s < \mu_s / \mu_b $, $S$ and $R_H$ have opposite signs (Region II). For 3D TIs, $S_s$ of the metallic surface states is much smaller than $S_b$ of the semiconducting bulk states (i.e., $|S_s / S_b| \ll 1$); $\mu_s$ of the surface states that are topologically protected is much higher than $\mu_b$ of the bulk states(i.e., $\mu_s / \mu_b \gg 1$). Therefore, there exists a sizable Region II, where experiments are able to observe opposite signs between $S$ and $R_H$. In this region, the sign of $S$ is determined by the states with stronger Seebeck effect, that are the bulk states with P-type Seebeck effect; the sign of $R_H$ is determined by the states with higher mobility, that is the surface states with N-type Hall effect. In fact, the parameter $\sigma_b / \sigma_s$ can be tuned by varying the chemical composition $x$ of the compound (Bi$_{1-x}$Sb$_x$)$_2$Te$_3$ and the temperature. When the temperature and/or $x$ increase, $\sigma_b / \sigma_s$ gradually increases. A transition from Region I to Region III would happen as predicted by theory [see Fig.~4(c)], which agrees well with the experimental observation [see Fig.~4(d)].

In the other experiment, we measured TE properties of 3D TI Bi$_2$Se$_3$ thin films from 5 QLs to 30 QLs, and observed strong size effects of TE quantities~\ucite{guo2016}. When decreasing the film thickness, the electrical resistance increases and the Seebeck coefficient decreases, resulting in decreased power factors. This is because for the thinner films, the relative contribution of the surface states gets more and more important. As the metallic surface states give weak Seebeck effect, the total $S$ gets suppressed by decreasing the film thickness. In this sense, the metallic surface states, if their advantage of high mobility is not fully utilized, would be detrimental to TE performance. In present experiments, $E_F$ is well within the bulk bands (caused by the large concentration of intrinsic dopants), which might lead to a low mobility of the surface states due to the strong surface-bulk interactions. Tuning the position of $E_F$  may significantly improve TE properties. Furthermore, the influence of the surface states can be studied by introducing a topological phase transition, for instance, by alloying, applying strain or tuning orbital levels~\ucite{liu2014,shi2015}.

\section{Outlook}
Most previous experiments on 3D TIs observed that the metallic surface states do not help improve TE performance due to their weak Seebeck effect. However, the great advantage of topological protection that ensures much weaker defect/disorder scattering on electrons of the surface states than on phonons has not been fully employed. Very recently, a TE experiment reported a record high $zT \sim 1.86$ for the P-type 3D TI material Bi$_{0.5}$Sb$_{1.5}$Te$_3$ at 320 K~\ucite{kim2015}. This is a significant breakthrough of TE research, considering that the $zT$ value is much improved over the previous record of $zT \sim 1$ for the same kind of materials. Topological states might play an important role in the enhancement of $zT$. Instead of using usual material samples, this work milled the bulk ingots of Bi$_{0.5}$Sb$_{1.5}$Te$_3$ into nanoparticles and then applied a new way of liquid-phase compaction to treat the sample. The treatment generates a lot of low-energy grain boundaries and dense dislocation arrays in the material. As a result, phonons are significantly scattered by the grain boundary and dislocation, especially the middle-frequency phonons that have a major contribution to thermal conductivity. The lattice thermal conductivity is thus strongly suppressed. On the other hand, the power factor of the material remains almost unchanged. $zT$ is greatly improved by this unusual feature that the grain boundary and dislocation induce strong scattering on phonons but not on electrons. However, the underlying mechanism is still unclear. What is role of the topological states in improving $zT$? Can we apply the technique to  enhance $zT$ of other materials? These are important open questions to be studied.

Current TE experiments mainly study 3D TI materials. In contrast, little experimental work has been done on TE properties of 2D TIs that are proposed to have enhanced $zT$~\ucite{xu2014}. Unfortunately, the existing 2D TIs (like HgTe/CdTe quantum well) have very low working temperatures (lower than 10 K). Large-gap 2D TI materials are still lacking for room temperature applications. In 2013 new 2D materials, stanene and its derivatives, were predicted by the first-principles calculations to be 2D TIs with sizeable band gaps up to 0.3 eV~\ucite{xu2013,tang2014}. So are the germanium counterparts, germanene and its derivatives~\ucite{si2014}. These materials as well as graphene systems with enhanced SOC~\ucite{li2012,li2013} are interesting due to the low dimensionality and the nontrivial band topology, which provide opportunities to explore new emerging physics and applications, like novel topological superconductivity~\ucite{wang2014} and near room temperature QAH effect~\ucite{wu2014}. As a very recent progress, the monolayer stanene structure has been successfully fabricated for the first time by molecular beam epitaxy and confirmed by scanning tunneling microscopy, ARPES together with first-principles calculations~\ucite{zhu2015}. Unfortunately, the obtained stanene is metallic due to the influence of the substrate. If choosing a suitable substrate, like SrTe(111)-Te, it is possible to get large-gap QSH states in decorated stanene~\ucite{xu2015}. Many following experiments are ongoing to grow the large-gap 2D TI materials and use them for TE research. Moreover, it is  interesting to investigate TE properties of other topological materials (like topological crystalline insulators and QAH insulators). These preliminary works may stimulate more TE research on topological materials. The continued work in this direction hopefully will discover more and more exotic TE effects and lead to significant breakthroughs in the improvement of TE efficiency.

\begin{center}
\includegraphics[width=0.8\textwidth]{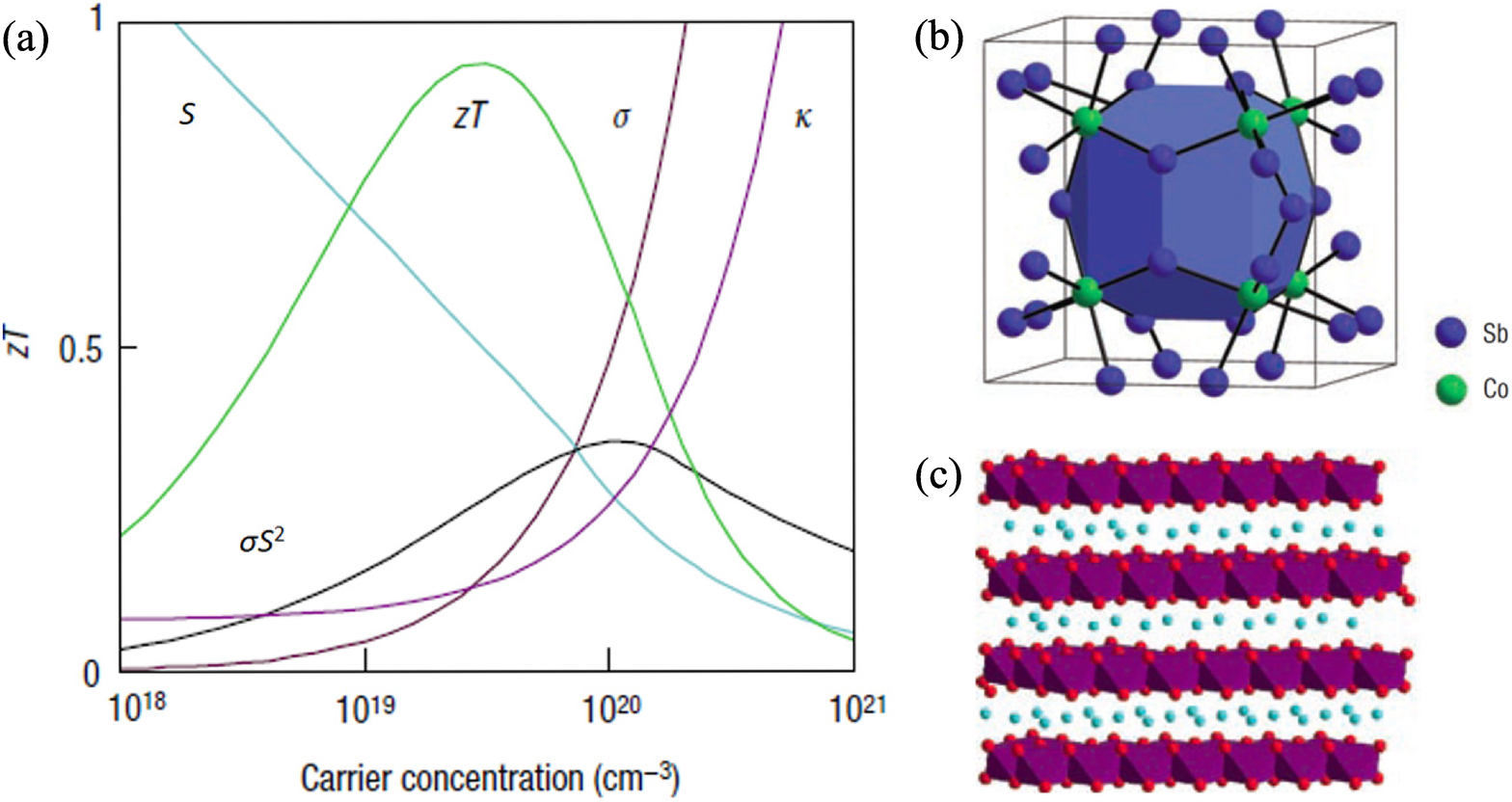}\\[5pt]
\parbox[c]{15.0cm}{\footnotesize{\bf Fig.~1.} (color online) (a) A schematic diagram showing the dependence of thermoelectric properties on carrier concentration. (b) The skutterudite CoSb$_3$ structure with large void space shown in blue. (c) Atomic structure of Na$_x$CoO$_2$ containing ordered layers (polyhedra) separated by disordered cation monolayers. Adapted from Ref.~\cite{snyder2008}.}
\end{center}

\begin{center}
\includegraphics[width=0.5\textwidth]{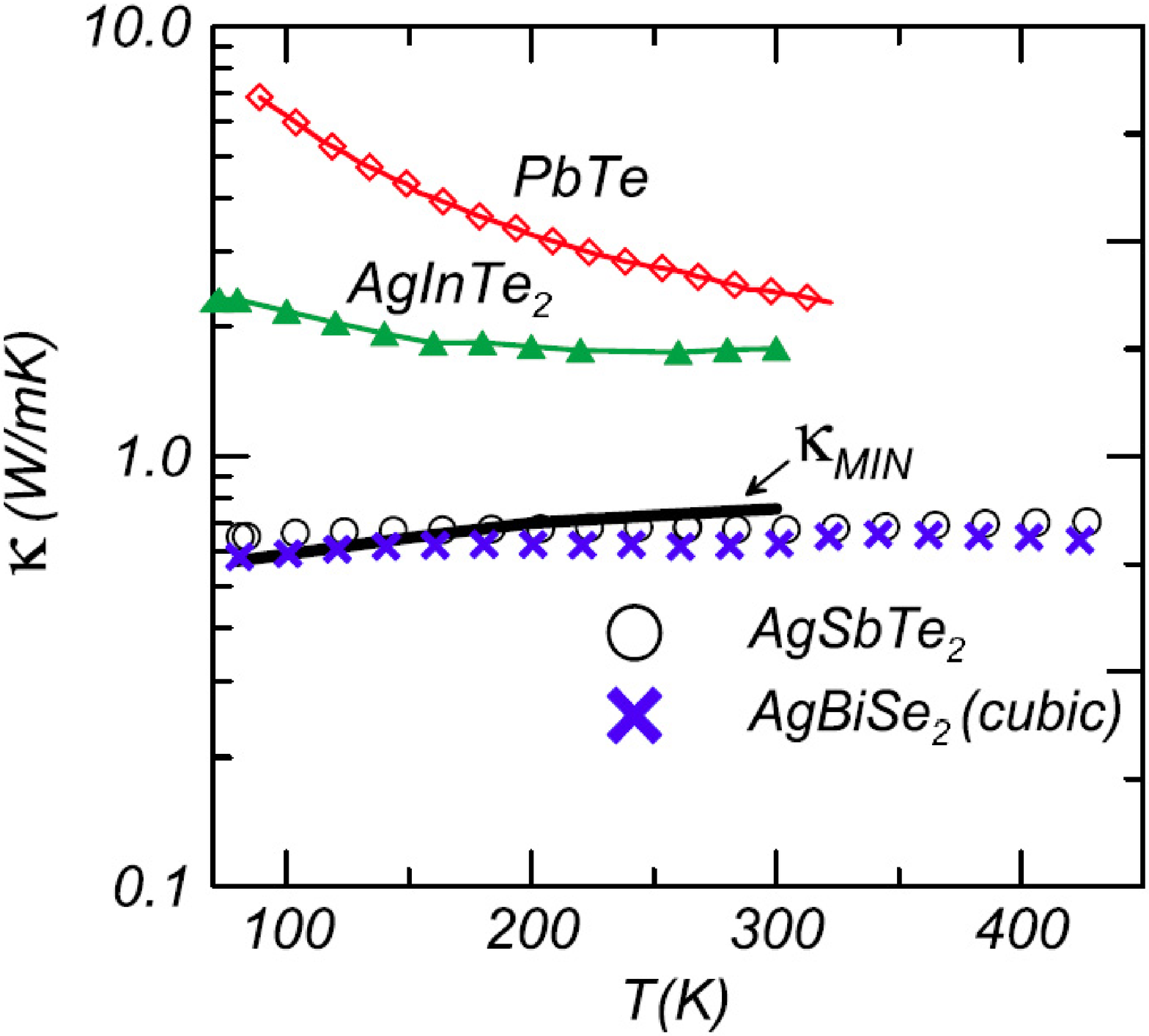}\\[5pt]
\parbox[c]{15.0cm}{\footnotesize{\bf Fig.~2.} (color online) The lattice thermal conductivity $\kappa$ as a function of temperature $T$. The black curve denote the calculated minimum thermal conductivity for AgSbTe$_2$. Data points are from experimental measurements. Adapted from Ref.~\cite{morelli2008}.}
\end{center}

\begin{center}
\includegraphics[width=\textwidth]{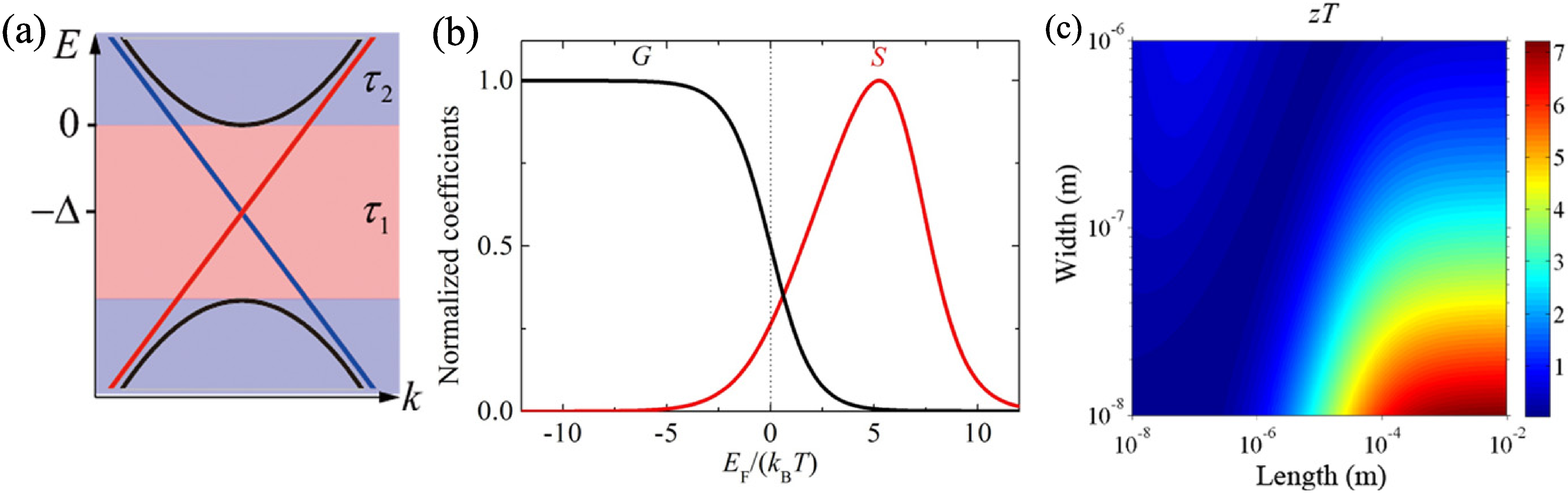}\\[5pt]
\parbox[c]{15.0cm}{\footnotesize{\bf Fig.~3.} (color online) (a) Schematic band structure of 2D TIs. $\tau_1$ and $\tau_2$ denote the scattering times within and outside the bulk-gap region, respectively. $\tau_1/\tau_2 \gg 1$. Red (blue) colored lines represent the edge states of up (down) spins. (b) The Fermi level $E_F$ dependence of TE properties for the edge states. $E_F$ is referenced to the bottom of the bulk conduction band. (c) The dependence of room-temperature $zT$ on the geometric parameters of nanorribon structures for the 2D TI material, fluorinated stanene. Adapted from Ref.~\cite{xu2014}.}
\end{center}

\begin{center}
\includegraphics[width=\textwidth]{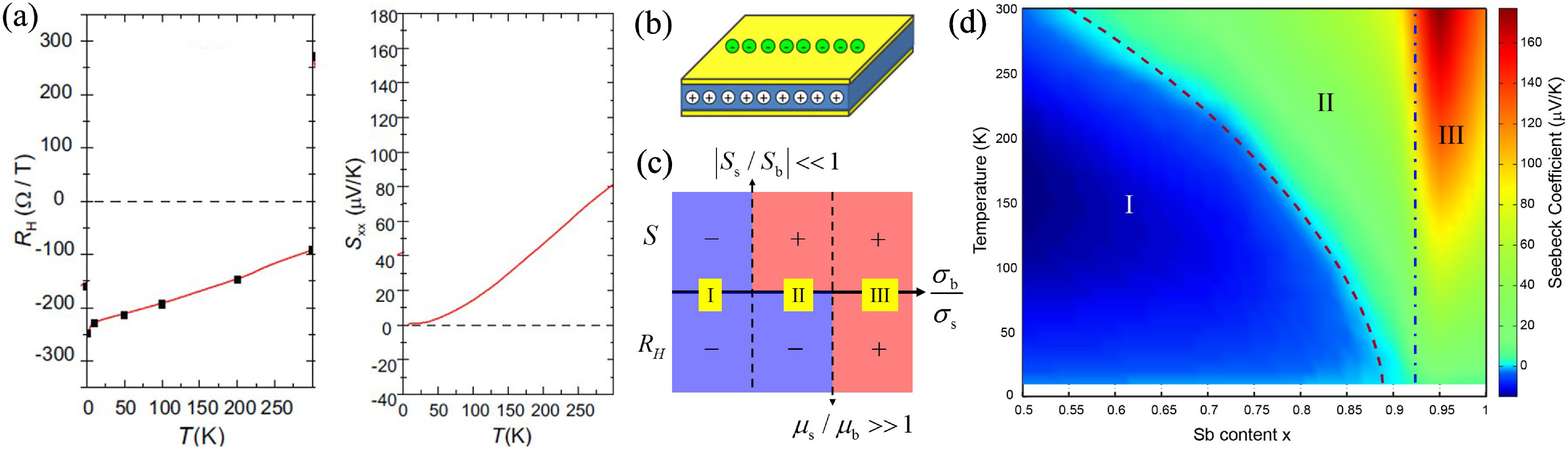}\\[5pt]
\parbox[c]{15.0cm}{\footnotesize{\bf Fig.~4.} (color online) TE properties of 3D TI (Bi$_{1-x}$Sb$_x$)$_2$Te$_3$ thin films. (a) The measured Hall (left) and Seebeck coefficients (right) as a function of temperature $T$ for the 5-quintuple-layer thin film with $x=0.9$. (b) Schematic drawing shows the coexistence of N-type surface states and P-type bulk states in a TI thin film. (c) A theoretical phase diagram for TI thin films with N-type surface and P-type bulk charge carriers. (d) The experimental phase diagram of (Bi$_{1-x}$Sb$_x$)$_2$Te$_3$ thin films summarizing the Seebeck coefficient as a function of $x$ and $T$. Three different regions (I,II,III) are defined according to the signs of $S$ and $R_H$ in (c,d). For TIs, there is a sizable region II that gives opposite signs between $S$ and $R_H$. Adapted from Ref.~\cite{zhang2015}.}
\end{center}

%\bibliography{references}

\begin{thebibliography}{10}
\expandafter\ifx\csname url\endcsname\relax
  \def\url#1{{\tt #1}}\fi
\expandafter\ifx\csname urlprefix\endcsname\relax\def\urlprefix{URL }\fi
\providecommand{\eprint}[2][]{\url{#2}}

\bibitem{hochbaum2008}
Hochbaum A~I, Chen R, Delgado R~D, Liang W, Garnett E~C, Najarian M, Majumdar A
  and Yang P 2008 {\em Nature\/} {\bf 451} 163

\bibitem{bell2008}
Bell L~E 2008 {\em Science\/} {\bf 321} 1457

\bibitem{snyder2008}
Snyder G~J and Toberer E~S 2008 {\em Nature Mater.\/} {\bf 7} 105

\bibitem{ioffe1957}
Ioffe A~F 1957 {\em Semiconductor thermoelements, and Thermoelectric cooling\/}
  (Infosearch, ltd.)

\bibitem{rowe1995}
Rowe D~M 1995 {\em CRC handbook of thermoelectrics\/} (CRC press)

\bibitem{goldsmid2010}
Goldsmid H~J 2010 {\em Introduction to thermoelectricity\/} vol 121 (Springer)

\bibitem{goldsmid1954}
Goldsmid H~J and Douglas R~W 1954 {\em British J. Appl. Phys.\/} {\bf 5} 386

\bibitem{hicks1993}
Hicks L and Dresselhaus M 1993 {\em Phys. Rev. B\/} {\bf 47} 12727

\bibitem{hicks1993_2}
Hicks L and Dresselhaus M 1993 {\em Phys. Rev. B\/} {\bf 47} 16631

\bibitem{chen2010}
Chen J, Zhang G and Li B 2010 {\em Nano Lett.\/} {\bf 10} 3978

\bibitem{chen2011}
Chen J, Zhang G and Li B 2011 {\em J. Chem. Phys.\/} {\bf 135} 204705

\bibitem{chen2009}
Chen J, Zhang G and Li B 2009 {\em Appl. Phys. Lett.\/} {\bf 95} 073117

\bibitem{shi2010}
Shi L, Yao D, Zhang G and Li B 2010 {\em Appl. Phys. Lett.\/} {\bf 96} 173108

\bibitem{yu2010}
Yu J~K, Mitrovic S, Tham D, Varghese J and Heath J~R 2010 {\em Nat.
  Nanotechnol.\/} {\bf 5} 718

\bibitem{tang2010}
Tang J, Wang H~T, Lee D~H, Fardy M, Huo Z, Russell T~P and Yang P 2010 {\em
  Nano Lett.\/} {\bf 10} 4279

\bibitem{chen2011_2}
Chen J, Zhang G and Li B 2011 {\em J. Chem. Phys.\/} {\bf 135} 104508

\bibitem{chen2012}
Chen J, Zhang G and Li B 2012 {\em Nano Lett.\/} {\bf 12} 2826

\bibitem{wingert2011}
Wingert M~C, Chen Z~C, Dechaumphai E, Moon J, Kim J~H, Xiang J and Chen R 2011
  {\em Nano Lett.\/} {\bf 11} 5507

\bibitem{morelli2008}
Morelli D, Jovovic V and Heremans J 2008 {\em Phys. Rev, Lett.\/} {\bf 101}
  035901

\bibitem{zhao2014}
Zhao L~D, Lo S~H, Zhang Y, Sun H, Tan G, Uher C, Wolverton C, Dravid V~P and
  Kanatzidis M~G 2014 {\em Nature\/} {\bf 508} 373

\bibitem{zhao2016}
Zhao L~D, Tan G, Hao S, He J, Pei Y, Chi H, Wang H, Gong S, Xu H, Dravid V~P
  {\em et~al.\/} 2016 {\em Science\/} {\bf 351} 141

\bibitem{kane2005}
Kane C~L and Mele E~J 2005 {\em Phys. Rev, Lett.\/} {\bf 95} 226801

\bibitem{bernevig2006}
Bernevig B~A, Hughes T~L and Zhang S~C 2006 {\em Science\/} {\bf 314} 1757

\bibitem{konig2007}
K{\"o}nig M, Wiedmann S, Br{\"u}ne C, Roth A, Buhmann H, Molenkamp L~W, Qi X~L
  and Zhang S~C 2007 {\em Science\/} {\bf 318} 766

\bibitem{qi2010}
Qi X~L and Zhang S~C 2010 {\em Phys. Today\/} {\bf 63} 33

\bibitem{hasan2010}
Hasan M~Z and Kane C~L 2010 {\em Rev. Mod. Phys.\/} {\bf 82} 3045

\bibitem{qi2011}
Qi X~L and Zhang S~C 2011 {\em Rev. Mod. Phys.\/} {\bf 83} 1057

\bibitem{Zhang2009}
Zhang H, Liu C~X, Qi X~L, Dai X, Fang Z and Zhang S~C 2009 {\em Nature Phys.\/}
  {\bf 5} 438--442

\bibitem{wang2013}
Wang J, Lian B, Zhang H, Xu Y and Zhang S~C 2013 {\em Phys. Rev. Lett.\/} {\bf
  111} 136801

\bibitem{zhang2014}
Zhang H, Xu Y, Wang J, Chang K and Zhang S~C 2014 {\em Phys. Rev. Lett.\/} {\bf
  112} 216803

\bibitem{zhang2014_2}
Zhang H, Wang J, Xu G, Xu Y and Zhang S~C 2014 {\em Phys. Rev. Lett.\/} {\bf
  112} 096804

\bibitem{zhao2015}
Zhao L, Wang J, Liu J, Xu Y, Gu B~L, Xue Q~K and Duan W 2015 {\em Phys. Rev.
  B\/} {\bf 92} 041408

\bibitem{xu2014}
Xu Y, Gan Z and Zhang S~C 2014 {\em Phys. Rev. Lett.\/} {\bf 112} 226801

\bibitem{ghaemi2010}
Ghaemi P, Mong R~S and Moore J~E 2010 {\em Phys. Rev. Lett.\/} {\bf 105} 166603

\bibitem{xu2009}
Xu Y, Chen X, Gu B~L and Duan W 2009 {\em Appl. Phys. Lett.\/} {\bf 95} 233116

\bibitem{xu2010}
Xu Y, Chen X, Wang J~S, Gu B~L and Duan W 2010 {\em Phys. Rev. B\/} {\bf 81}
  195425

\bibitem{zhu2012}
Zhu H, Xu Y, Gu B~L and Duan W 2012 {\em New J. Phys.\/} {\bf 14} 013053

\bibitem{chen2013}
Chen X, Xu Y, Zou X, Gu B~L and Duan W 2013 {\em Phys. Rev. B\/} {\bf 87}
  155438

\bibitem{huang2013}
Huang H, Xu Y, Zou X, Wu J and Duan W 2013 {\em Phys. Rev. B\/} {\bf 87} 205415

\bibitem{li2014}
Li D, Xu Y, Chen X, Li B and Duan W 2014 {\em Appl. Phys. Lett.\/} {\bf 104}
  143108

\bibitem{zou2015}
Zou X, Chen X, Huang H, Xu Y and Duan W 2015 {\em Nanoscale\/} {\bf 7} 8776

\bibitem{chen2015}
Chen X~B and Duan W 2015 {\em Acta Phys. Sin.\/} {\bf 64} 186302

\bibitem{paulsson2003}
Paulsson M and Datta S 2003 {\em Phys. Rev. B\/} {\bf 67} 241403

\bibitem{xu2008}
Xu Y, Wang J~S, Duan W, Gu B~L and Li B 2008 {\em Phys. Rev. B\/} {\bf 78}
  224303

\bibitem{xu2014small}
Xu Y, Li Z and Duan W 2014 {\em Small\/} {\bf 10} 2182

\bibitem{takahashi2010}
Takahashi R and Murakami S 2010 {\em Phys. Rev. B\/} {\bf 81} 161302

\bibitem{murakami2011}
Murakami S, Takahashi R, Tretiakov O, Abanov A and Sinova J 2011 {\em J. Phys.:
  Conf. Ser.\/} {\bf 334} 012013

\bibitem{takahashi2012}
Takahashi R and Murakami S 2012 {\em Semicond. Sci. Technol.\/} {\bf 27} 124005

\bibitem{chang2014}
Chang P~H, Bahramy M~S, Nagaosa N and Nikolic B~K 2014 {\em Nano Lett.\/} {\bf
  14} 3779

\bibitem{xu2013}
Xu Y, Yan B, Zhang H~J, Wang J, Xu G, Tang P, Duan W and Zhang S~C 2013 {\em
  Phys. Rev. Lett.\/} {\bf 111} 136804

\bibitem{venkatasubramanian2001}
Venkatasubramanian R, Siivola E, Colpitts T and O'quinn B 2001 {\em Nature\/}
  {\bf 413} 597

\bibitem{huang2008}
Huang B~L and Kaviany M 2008 {\em Phys. Rev. B\/} {\bf 77} 125209

\bibitem{hor2009}
Hor Y~S, Richardella A, Roushan P, Xia Y, Checkelsky J~G, Yazdani A, Hasan M~Z,
  Ong N~P and Cava R~J 2009 {\em Phys. Rev. B\/} {\bf 79} 195208

\bibitem{zahid2010}
Zahid F and Lake R 2010 {\em Appl. Phys. Lett.\/} {\bf 97} 212102

\bibitem{yu2014}
Yu C, Zhang G, Peng L~M, Duan W and Zhang Y~W 2014 {\em Appl. Phys. Lett.\/}
  {\bf 105} 023903

\bibitem{rittweger2014}
Rittweger F, Hinsche N~F, Zahn P and Mertig I 2014 {\em Phys. Rev. B\/} {\bf
  89} 035439

\bibitem{osterhage2014}
Osterhage H, Gooth J, Hamdou B, Gwozdz P, Zierold R and Nielsch K 2014 {\em
  Appl. Phys. Lett.\/} {\bf 105} 123117

\bibitem{gooth2014}
Gooth J, Gluschke J~G, Zierold R, Leijnse M, Linke H and Nielsch K 2014 {\em
  Semicond. Sci. Technol.\/} {\bf 30} 015015

\bibitem{yu2015}
Yu C, Zhang G, Zhang Y~W and Peng L~M 2015 {\em Nano Energy\/} {\bf 17} 1040

\bibitem{hinsche2015}
Hinsche N~F, Zastrow S, Gooth J, Pudewill L, Zierold R, Rittweger F, Rauch T,
  Henk J, Nielsch K and Mertig I 2015 {\em ACS Nano\/} {\bf 9} 4406

\bibitem{shi2015_2}
Shi H, Parker D, Du M~H and Singh D~J 2015 {\em Phys. Rev. Appl.\/} {\bf 3}
  014004

\bibitem{liu2016}
Liu W, Chi H, Walrath J, Chang A, Stoica V~A, Endicott L, Tang X, Goldman R and
  Uher C 2016 {\em Appl. Phys. Lett.\/} {\bf 108} 043902

\bibitem{liang2016}
Liang J, Cheng L, Zhang J, Liu H and Zhang Z 2016 {\em Nanoscale\/} {\bf 8}
  8855

\bibitem{zhang2015}
Zhang J, Feng X, Xu Y, Guo M, Zhang Z, Ou Y, Feng Y, Li K, Zhang H, Wang L {\em
  et~al.\/} 2015 {\em Phys. Rev. B\/} {\bf 91} 075431

\bibitem{guo2016}
Guo M, Wang Z, Xu Y, Huang H, Zang Y, Liu C, Duan W, Gan Z, Zhang S~C, He K
  {\em et~al.\/} 2016 {\em New J. Phys.\/} {\bf 18} 015008

\bibitem{liu2014}
Liu J, Xu Y, Wu J, Gu B~L, Zhang S and Duan W 2014 {\em Acta Cryst. C\/} {\bf
  70} 118

\bibitem{shi2015}
Shi W~J, Liu J, Xu Y, Xiong S~J, Wu J and Duan W 2015 {\em Phys. Rev. B\/} {\bf
  92} 205118

\bibitem{kim2015}
Kim S~I, Lee K~H, Mun H~A, Kim H~S, Hwang S~W, Roh J~W, Yang D~J, Shin W~H, Li
  X~S, Lee Y~H {\em et~al.\/} 2015 {\em Science\/} {\bf 348} 109

\bibitem{tang2014}
Tang P, Chen P, Cao W, Huang H, Cahangirov S, Xian L, Xu Y, Zhang S~C, Duan W
  and Rubio A 2014 {\em PPhys. Rev. B\/} {\bf 90} 121408

\bibitem{si2014}
Si C, Liu J, Xu Y, Wu J, Gu B~L and Duan W 2014 {\em Phys. Rev. B\/} {\bf 89}
  115429

\bibitem{li2012}
Li Y, Chen P, Zhou G, Li J, Wu J, Gu B~L, Zhang S and Duan W 2012 {\em Phys.
  Rev. Lett.\/} {\bf 109} 206802

\bibitem{li2013}
Li Y, Tang P, Chen P, Wu J, Gu B~L, Fang Y, Zhang S and Duan W 2013 {\em Phys.
  Rev. B\/} {\bf 87} 245127

\bibitem{wang2014}
Wang J, Xu Y and Zhang S~C 2014 {\em Phys. Rev. B\/} {\bf 90} 054503

\bibitem{wu2014}
Wu S~C, Shan G and Yan B 2014 {\em Phys. Rev. Lett.\/} {\bf 113} 256401

\bibitem{zhu2015}
Zhu F~F, Chen W~J, Xu Y, Gao C~L, Guan D~D, Liu C~H, Qian D, Zhang S~C and Jia
  J~F 2015 {\em Nature Mater.\/} {\bf 14} 1020

\bibitem{xu2015}
Xu Y, Tang P and Zhang S~C 2015 {\em Phys. Rev. B\/} {\bf 92} 081112

\end{thebibliography}

\end{CJK*}  %% end the Chinese environment
\end{document}